\documentclass[sigconf]{acmart}
\AtBeginDocument{%
  }

\copyrightyear{2023}
\acmYear{2023}
\setcopyright{none}

\acmConference[Conference acronym]{Conference acronym}{2023}{Preprint}
\acmPrice{15.00}
\acmDOI{10.1145/nnnnnnn.nnnnnnn}
\acmISBN{979-x-xxxx-xxxx-x/YY/MM}

\usepackage[mathlines]{lineno}
\usepackage{mathtools}
\usepackage{multirow}
\usepackage{nccmath}
\usepackage{tabularx}
\usepackage{graphicx}
\usepackage{booktabs}
\usepackage{enumitem}
\usepackage{tablefootnote}
\usepackage{color}
\usepackage{soul}
\usepackage{physics}
\usepackage{subcaption}
\usepackage{xcolor}

\usepackage{multicol}
\begin{document}

\title{
%Learning to Assign socioeconomic Indicators by Observing Inequality through Satellite Images
%A Weak Supervision Method to Infer Socioeconomic Indicators from Satellite Imagery
%Seeing Beyond the Surface: Inferring Finer Scale Socioeconomic Indicators from Satellite Imagery
Fine-Grained Socioeconomic Prediction from Satellite Images with Distributional Adjustment}
%\author{Anonymous authors}

\author{Donghyun Ahn}
\orcid{0000-0001-8473-9598}
\affiliation{%
 \institution{KAIST, IBS}
 \city{Daejeon}
 \country{South Korea}
}
\email{segaukwa@kaist.ac.kr}

\author{Minhyuk Song}
\orcid{0009-0004-3634-2560}
\affiliation{%
 \institution{KAIST, IBS}
 \city{Daejeon}
 \country{South Korea}
}
\email{smh0706@kaist.ac.kr}

\author{Seungeon Lee}
\orcid{0000-0002-9756-0068}
\affiliation{%
 \institution{KAIST, IBS}
 \city{Daejeon}
 \country{South Korea}
}
\email{archon159@kaist.ac.kr}

\author{Yubin Choi}
\orcid{0000-0002-6042-5313}
\affiliation{%
 \institution{KAIST, IBS}
 \city{Daejeon}
 \country{South Korea}
}
\email{yubin.choi@kaist.ac.kr}

\author{Jihee Kim}
\orcid{0000-0002-3875-1083}
\affiliation{%
 \institution{KAIST, IBS}
 \city{Daejeon}
 \country{South Korea}
}
\email{jiheekim@kaist.ac.kr}

\author{Sangyoon Park}
\orcid{0000-0003-3078-5956}
\affiliation{%
\institution{HKUST}
  \city{Hong Kong}
  \country{China}
}
\email{sangyoon@ust.hk}

\author{Hyunjoo Yang}
\orcid{0000-0001-7204-5919}
\affiliation{%
  \institution{Sogang University}
  \city{Seoul}
  \country{South Korea}
}
\email{hyang@sogang.ac.kr}

\author{Meeyoung Cha}
\orcid{0000-0003-4085-9648}
\affiliation{%
 \institution{IBS, KAIST}
 \city{Daejeon}
 \country{South Korea}
}
\email{meeyoungcha@kaist.ac.kr}

  %\institution{Institute for Basic Science}
  %\city{Daejeon}
  %\country{Republic of Korea}
%\affiliation{
  %\institution{Korea Advanced Institute of Science and Technology}
  %\city{Daejeon}
  %\country{Republic of Korea}
%}

\renewcommand{\shortauthors}{Donghyun Ahn, et al.}
%\email{webmaster@marysville-ohio.com}


\begin{abstract}
While measuring socioeconomic indicators is critical for local governments to make informed policy decisions, such measurements are often unavailable at fine-grained levels like municipality. This study employs deep learning-based predictions from satellite images to close the gap. We propose a method that assigns a socioeconomic score to each satellite image by capturing the distributional behavior observed in larger areas based on the ground truth. We train an ordinal regression scoring model and adjust the scores to follow the common power law within and across regions. Evaluation based on official statistics in South Korea shows that our method outperforms previous models in predicting population and employment size at both the municipality and grid levels. Our method also demonstrates robust performance in districts with uneven development, suggesting its potential use in developing countries where reliable, fine-grained data is scarce.
\end{abstract}



\newcommand{\img}{{\bf x}}
\newcommand{\municipality}{\mathcal{M}}
\newcommand{\district}{\mathcal{D}}
\newcommand{\ctry}{\mathcal{C}}
\newcommand{\mc}[1]{\textcolor{magenta}{#1}}
\newcommand{\tony}[1]{\textcolor{blue}{#1}}
\newcommand{\mh}[1]{\textcolor{brown}{#1}}
\newcommand{\yb}[1]{\textcolor{gray}{#1}}
\newcommand{\jihee}[1]{\textcolor{violet}{ #1}}
\newcommand{\sean}[1]{\textcolor{olive}{#1}}
\newcommand{\syp}[1]{\textcolor{red}{#1}}
\maketitle

\section{Introduction}
\begin{figure}[t!]
\centering
    \begin{subfigure}[h]{0.18\textwidth}
      \includegraphics[width=\textwidth]{source/Figs/Figure_1a_v2.jpg}
    \end{subfigure}
    \hspace{-0.5mm}
    \begin{subfigure}[h]{0.29\textwidth}
      \includegraphics[width=\textwidth]{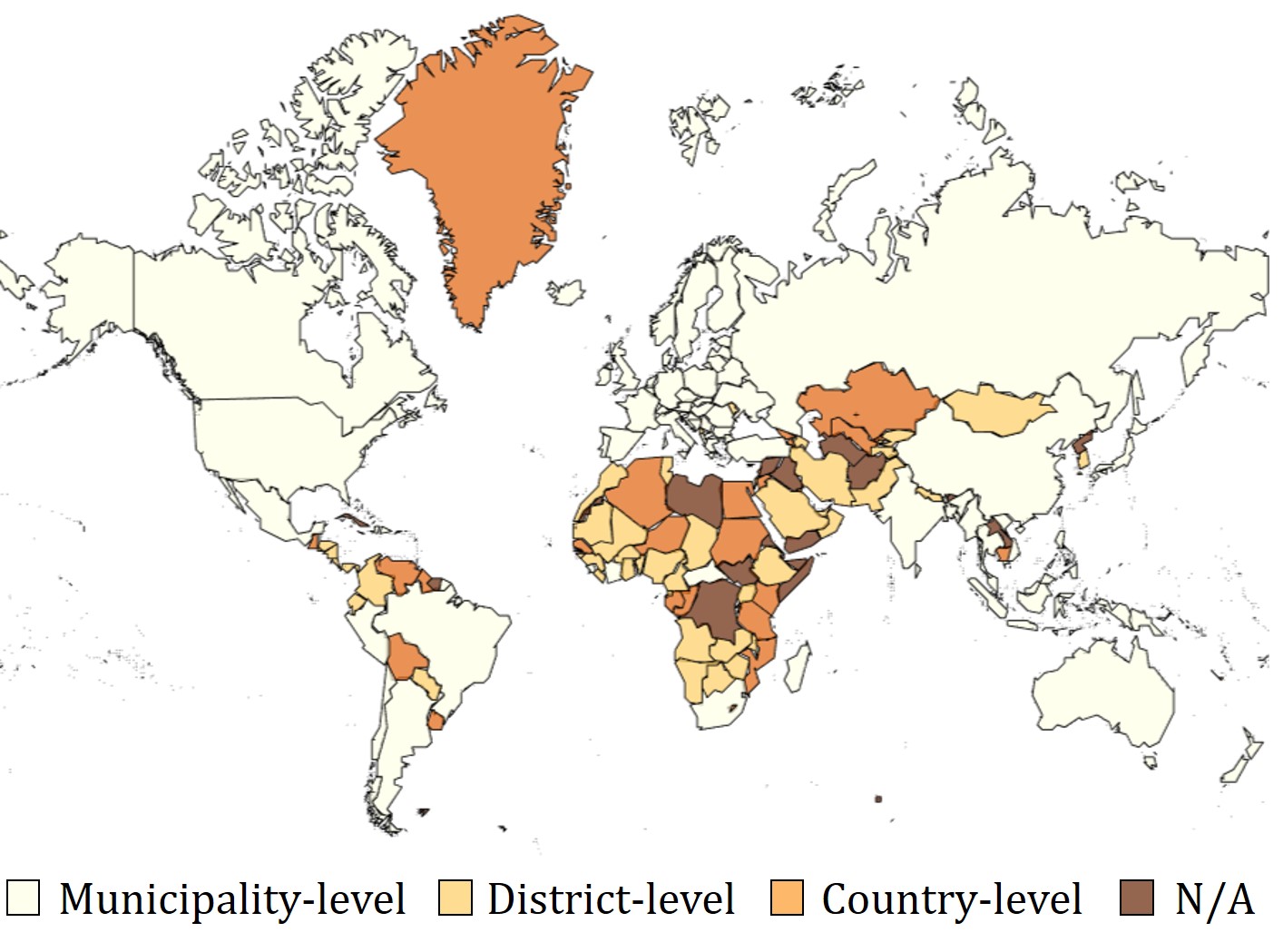}
    \end{subfigure}    
    %\hspace{0.5mm}
    \vspace*{-2mm}
\caption{The administrative level structure and a global map displaying the availability of Esri Demographics~\cite{esridemographics21} encompassing 170+ countries at three administrative levels: country, district, and municipality.}
\label{fig:concept}
\end{figure}

Many policy decisions can benefit from accurate measurements of socioeconomic statistics such as population, businesses, income, birth, and death rates across geography. Such information would enable local governments to make informed decisions and develop effective policies at a granular level that are tailored to their specific local conditions. One recent example was the distribution of COVID-19 vaccines, where local governments were able to prioritize seniors and healthcare workers~\cite{dooling2021advisory} by leveraging the knowledge of citizens' geographical distribution by age and occupation~\cite{krzysztofowicz2021use}. Vaccine allocation strategies were optimized by demographic and occupational characteristics at a granular level, ensuring that high-risk groups were prioritized.

However, producing these statistics at granular administrative units remains challenging, particularly in developing countries with limited resources. Figure~\ref{fig:concept} depicts the hierarchical and administrative structures based on Esri Demographics~\cite{esridemographics21}, a database for socioeconomic indicators at a sub-national level covering broad global regions. 
Only 13\% of the developing countries release municipality-level data, whereas 35 out of 50 (20\%) wealthiest countries own data at this level~\cite{egilitis2019worlddata}.
Furthermore, conventional data collection methods like population census are reliable but expensive and time-consuming, making them difficult to conduct frequently. 
 
As a result, alternative data sources, such as the credit card usage~\cite{oosterhof2018localizing} or street images~\cite{chetty2022social}, are increasingly being used to infer wealth at finer levels. Another data source is satellite imagery, which offers comprehensive geographic coverage~\cite{abitbol2020interpretable,yeh2020using,chi2022microestimates}. Several recent studies have applied computer vision techniques to estimate socioeconomic indicators over large areas~\cite{jean2016combining,han2020lightweight,han2020learning} using satellite images. The inference of wealth (or poverty) over smaller areas, however, remains a new frontier. 

We present a new method for estimating the socioeconomic status of small administrative units by assigning scores to individual satellite images. By aligning these scores with expected distribution statistics, our method refines the data aggregated over larger areas (e.g., districts) to a more granular scale. First, we utilize a weakly supervised ordinal regression to establish an initial scoring function across all satellite images, using a limited set of sample labels~\cite{park2022learning}. Next, our model learns the internal distribution within each district, aligning it with established theories. Specifically, we steer the scoring function so that aggregated scores across small administrative divisions (e.g., municipalities) correspond to the ratio implied by the common power law, an empirical economic theory with theoretical explanations~\cite{mori2020common}. Our technique enhances prediction performance and maintains the additive property, enabling more precise aggregation of estimated economic scores across neighboring regions. For evaluation, we test with publicly available socioeconomic indicators at the district level, which serve as ground truth.

\section{Datasets}
\subsection{Satellite imagery}

We utilized the World Imagery~\cite{esri2021world} service to access satellite images of South Korean regions provided by WorldView-2 and GeoEye. The satellite images were accessed via the ArcGIS Developer subscription. This study used 109,643 images with a resolution of 4.8 meters per pixel. Most images were taken between the spring and autumn seasons from 2018 to 2020 for visual consistency and clarity. We avoided images from winter because of the snow.
 
\subsection{Socioeconomic indicators}
We consider two indicators that adhere to the power law in estimating city size~\cite{gabaix2016power}:
\begin{itemize}[leftmargin=*]
  \item \textbf{Population} is a commonly used indicator to approximate the city size. This statistic is available for most countries at varying levels. The South Korean National Geographic Information Institute (NGII) provided the resident registration population for each 1 km by 1 km grid area in 2020. For this study, 75,334 grid-level population data points were utilized. We obtained the population of municipalities and districts by aggregating this grid-level data.
  \item \textbf{Employment size} is the total employment in a given area and is used as an alternative indicator of city size. We downloaded the official employment size data for 2020, released by Statistics Korea. This data includes the complete list of business establishments and their employment size across 2,260 municipalities.
\end{itemize}

\if 0
We used the World Imagery~\cite{esri2021world} data that is provided by WorldView-2 and GeoEye. The RGB band images in the dataset have a maximum spatial resolution of about 0.5m. We chose 256 by 256-pixel images with a 4.8m resolution to predict social activities within a square area ($\approx$1.44$km^2$). The ArcGIS Developer subscription was used to access satellite imagery collections for South Korea, with a total of 109,643 images used for this study. Most images were captured between 2018 and 2020 during the spring and autumn seasons to reduce potential noise from snow-laden imagery.
\fi

\if 0
Our scoring model requires district-level ground data to infer the scale of actual value, where the distribution of integrated score among municipalities within a district aligns with the common power law~\cite{mori2020common}. The previous research~\cite{gabaix2016power} has demonstrated that both population and employment sizes serve as representative socioeconomic indicators adhere to this empirical law in estimating city size. Although district data alone is sufficient to train the model, we collect as much finer data as possible for evaluating the prediction and analyzing the model's performance.
\fi
 
\section{Model}
\subsection {Overview}
Given a set of satellite images $\ctry = \{ \textbf{x}_1, ..., \textbf{x}_k \}$ that cover an entire country, we analyze $n$ districts denoted as $\district_1, \district_2, ... , \district_n$, where each district is a partition of $\ctry$. Every district $\district_i$ is again divided into $m_i$ municipalities $\municipality_{i_1}$, $\municipality_{i_2}$, ..., $\municipality_{i_{m_i}}$ establishing a partition within $\district_i$. Leveraging the fractal structure of our data (i.e., smaller components of the system resemble the larger ones) that partitions the entire country $\ctry$ and using the publicly available ground indicators for each $\district_i$, our goal is to train a score function $f : \ctry \rightarrow [0,\infty]$ that computes the estimated economic indicator $\hat{y}$ for a given input image $\img$. Notably, $f$ is designed to establish an ordinal relationship among the images while ensuring that the aggregated sum of $f$ across each municipality adheres to an adjustment distribution and maintains linear scalability with the respective ground truth.

Given the conditions above, our model proceeds in two steps:
\begin{itemize}[leftmargin=*]
    \item \textbf{Step 1. Initialize $f$ with weak supervision.} We train the initial scoring function $f_0$ using a weakly-supervised method~\cite{park2022learning}. 
    \item \textbf{Step 2. Distributionally adjust $f$ with power-law.} 
    Starting from $f_{0}$, we guide our scoring function $f$ to ensure that the sum of scores across municipalities conforms to the expected ratio, which is derived using the given distribution for adjustment. We employ the common power law.
\end{itemize}

The power law is an economic theory that explains the allocation of city sizes within a given region~\cite{mori2020common}, which we also use as the basis for establishing the adjusting distribution. Note that $f$ can be trained on a different scale. We calibrate the model further with the ground truth by minimizing the ratio difference between the total sum of $f$ and the ground truth across two districts.

\subsection{Method}

We explain the two steps in detail:\\
\textbf{Step 1. Initialize $f$ with weak supervision} \\
We first train the initial scoring function $f_0$, which is trained in a weakly supervised manner. A recent work~\cite{park2022learning} proposed a method for estimating the economic scale of single satellite imagery by minimizing the cross entropy between the surrogate sample soft labels and the logit vectors that come from the ordinal regression~\cite{brant1990assessing}. We consider three class labels, including urban, rural, and uninhabited. Following the existing literature, we chose a random set of 1,000 images $\mathcal{R} = \{ \textbf{x}_1, ..., \textbf{x}_{1000} \} \subset \ctry$, and five annotators contributed to labeling. This labeling task was completed in approximately 2 hours with a Fleiss' $\kappa=0.77$, indicating substantial agreement among the annotators. We then train the initial scoring function $f_0$ by following the method described in a previous study~\cite{park2022learning}, where the order of image scores aligns well with the actual economic scale. %\\

\begin{figure}[htpb!]
\centering
\includegraphics[width=0.5\textwidth]{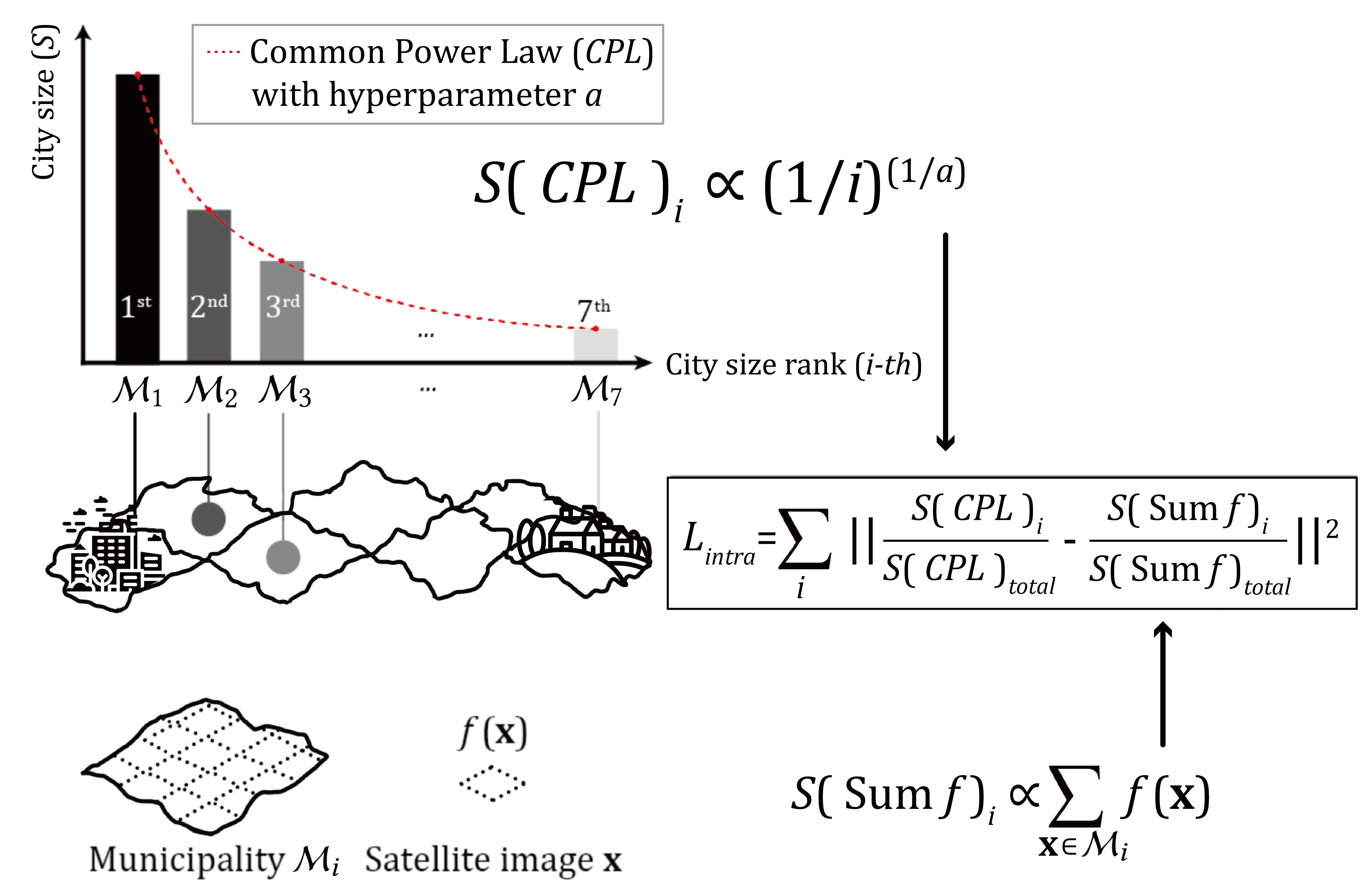} 
\vspace*{-6mm}
\caption{Illustration of the loss objective $L_{intra}$.}
\label{fig:L_intra}
\end{figure}

\vspace*{1mm}
\noindent
%\textbf{Step 2. Distributional adjustment on $f$} \\
\textbf{Step 2. Distributionally adjust $f$ with power-law}\\
Starting from the base function $f_0$, we can further improve the score model to learn the behavior of city size distribution inside each district. Previous studies have shown that city size distribution across various countries follows the power law property~\cite{gabaix2016power,gabaix2004evolution}. Recent economic studies further extend the scope of this phenomenon, arguing that it also encapsulates the distribution of cities within a country~\cite{mori2020common}. They provide evidence of spatial fractal structures that convey this power law even at a finer scale. The study points out that if the system has a fractal structure and follows power law, then the smaller parts also follow similar power laws. Focusing on the hierarchy, Figure~\ref{fig:L_intra} depicts our novel loss objective $L_{intra}$ that the cumulative sum of $f$ values across municipalities aligns with the common power law distribution within a district. 

The Pareto distribution with a shape variable $a$ drives the power law, implying the degree of economic inequality across the entire system~\cite{arnold2014pareto}. The common power law states in the formula that the probability of a city size $S$ greater than $s$ is given by:\useshortskip
\begin{linenomath}
\begin{align}
Pr(S>s) \approx cs^{-a} 
\label{eq:cpl_1}
\end{align}
\end{linenomath}
where $a$ comes from pareto distribution. Let, there are total $n$ municipalities $\municipality_1, \municipality_2, ..., \municipality_n$ in given district $\district$ and their city size (i.e., $S_{1}, S_{2}, ... S_{n}$) distribution is postulated to follow the common power law. Then, if these city size of municipalities are ranked as $S_1 \ge S_2 \ge ..., S_n$, so that $i$ implies the rank of $\municipality_i$,
we can obtain following approximation from Equation~\ref{eq:cpl_1}.\useshortskip
\begin{linenomath}
\begin{align}
i/n = Pr(S>s_{i}) \approx c{s_{i}}^{-a} \Rightarrow s_{i} \approx (nc/i)^{1/a}
\label{eq:cpl_2}
\end{align}
\end{linenomath}

% Decided not to change this part: WHY 7?
As our model learns the distribution of aggregated $f$ within the district, we set the minimum number of finer regions to seven to ensure that the distribution has enough data points to learn. This decision means that the model will ignore districts with six or fewer municipalities. We then aggregate $f$ over the top seven municipalities and define a loss objective ${L}_{intra}$ as the mean-squared error between the L1-normalized sum of $f$ across a municipality and the pseudo-label at the same location, informed by common power law as follows:\useshortskip
\begin{linenomath}
\begin{align}
{L}_{intra} =  \sum_{i=1}^m{\norm{\frac{\sum_{\img \in \municipality_{i}}{f(\img)}}{\sum_{j=1}^m {\sum_{\img \in \municipality_{j}}{f(\img)}}} -\frac{(1/i)^{1/a}}{\sum_{j=1}^m{(1/j)^{1/a}}}}^{2}} 
\label{eq:loss_fit}
\end{align}
\end{linenomath}
where $m$ denotes the top-$m$ municipalities of significant importance those are used to train the model (i.e., we set $m$ as 7).

The score $f$ undergoes further adjustment by taking into account the ratio between the aggregate of image scores within a district and the respective ground truth, across two different areas.
%linear relationship that enables the aggregation of estimated score $f$ over a given area.
Consequently, scoring function $f$ is also expected to follow:\useshortskip
\begin{linenomath*}
\begin{flalign}
G_{i} \propto {\sum\nolimits_{\img \in \mathcal{D}_{i}}{f(\img)}}, \frac{G_{i}}{G_{j}} \approx \frac{\sum_{\img \in \mathcal{D}_{i}}{f(\img)}}{\sum_{\img \in {\mathcal{D}_{j}}}{f(\img)}}
\label{eq:linear}
\end{flalign}
\end{linenomath*}
where $G_{i}$ represents the ground-truth indicator of district $\mathcal{D}_{i}$. This equation holds across the districts $\mathcal{D}_{i}$ and $\mathcal{D}_{j}$. To optimize the training process and reduce overhead, we selectively pick two districts $\mathcal{D}_{i-}$ and $\mathcal{D}_{i+}$ such that $G_{i-}<G_{i}<G_{i+}$, with the values being closest, respectively. Given that $\mathcal{D}_{i-}$ and $\mathcal{D}_{i+}$ possess city sizes similar to $\mathcal{D}_{i}$, this approach also prevents the proportion of the cumulative sum of $f$ across districts from diverging as much as possible. Utilizing Eqation~\ref{eq:linear}, we suggest the new loss objective ${L}_{inter}$, which is defined as follows:\useshortskip
\begin{linenomath}
\begin{align}
&{L}_{inter}={L}_{i-}+{L}_{i+}, \nonumber \\ 
&{L}_{i-}=\norm{\frac{G_{i-}}{G_{i}} - \frac{\sum_{\img \in \mathcal{D}_{i-}}{f(\img)}}{\sum_{\img \in {\mathcal{D}_{i}}}{f(\img)}}}^{2},
{L}_{i+}=\norm{\frac{G_{i}}{G_{i+}}-\frac{\sum_{\img \in \mathcal{D}_{i}}{f(\img)}}{\sum_{\img \in {\mathcal{D}_{i+}}}{f(\img)}}}^{2}
\end{align}
\label{eq:loss_adm2}
\end{linenomath}
\noindent
To integrate all the steps and ideas mentioned above, the score function $f$ is trained by optimizing the composite loss objective ${L}$:
\begin{linenomath}
\begin{align}
{L} = {L}_{intra} + {L}_{inter} 
\label{eq:loss_fit}
\end{align}
\end{linenomath}

\section{Evaluations and Analyses}
\subsection{Implementation Settings}
We utilize Resnet-18~\cite{he2016deep} as the backbone and trained for 100 epochs with batch size 50 for initializing $f$. 
Hyperparmeter $a$ is an shape variable for the power law that reveals the economic inequality of the given country. We tested for $a= log_{4}5, log_{9}10$, and $1$, where each explains 80-20, 90-10, and 100-0 rule according to the pareto principle~\cite{dunford2014pareto}, respectively. We empirically select $a= log_{9}10$ for South Korea. The codes and implementation details can be found in the project repository\footnote{\url{https://github.com/archive-cs-minhyuk/Distribution-learning}}.

\subsection{Performance evaluation}
We used several baselines for evaluation. All evaluations were performed on a South Korean dataset using an 80/20 train/test split, and the prediction model was evaluated for population and employment size. Because the function $f$ is trained to have a score proportional to the real quantity, we used R-squared $(R^2)$ at two different scales: the district level and the country level.
For the district level, we reported the average $R^{2}$ across municipalities and grids within each district.  
We used all the data in the test dataset at the country level. This setting is in accordance with the widely adopted evaluation method in this field~\cite{park2022learning,xi2022beyond}.

We applied L1-normalization to the grid-level score predictions to ensure that the sum of scores across the entire district aligns with the known district-level ground data. For a fair comparison, we normalized the results from other baseline models in the same way. We also evaluated the prediction at the municipality level, which can measure the performance regarding how well the predicted values can be aggregated. These two settings allow us to examine whether our model predicts a more accurate value without overestimating the model's performance. Illustrated in Table~\ref{tab:tbmain}, we compared our model with four baselines introduced as follows:

\begin{table}[ht!]
\centering
\caption{Comparison of prediction models over South Korea, highlighting the best (bolded) and the second best (underlined) performances. Results are evaluated by $R^{2}$ for population (Pop) and employment size (Employ) at the municipality and grid levels and across districts and a country.}
\label{tab:tbmain}
\makebox[0.4\textwidth][c]{
\scalebox{0.75}{
\begin{tabular}{lcccc|c}
\toprule
\multicolumn{1}{c}{\multirow{2}{*}{Models}} & \multicolumn{2}{c|}{Population (district)} & \multicolumn{2}{c|}{Population (country)} & \multicolumn{1}{c}{Employ (district)} \\ \cmidrule{2-6}
\multicolumn{1}{c}{}  & Municipality & \multicolumn{1}{c|}{Grid} & Municipality & Grid & Municipality   \\ \midrule
Resnet-18~\cite{he2016deep}                                   & -0.0410   & \multicolumn{1}{c|}{0.3831}  & 0.9060 & 0.6964 & -        \\
AutoEncoder~\cite{kramer1991nonlinear}                                 & -0.2087   & \multicolumn{1}{c|}{-0.0250}  & 0.8793 & 0.4393 & -       \\
PCA~\cite{tipping1999probabilistic}
 & -1.1900   & \multicolumn{1}{c|}{0.0920}  & 0.7442 & 0.2648 & -       \\
POI~\cite{xi2022beyond}                                         & 0.6472 & \multicolumn{1}{c|}{0.5760}& \textbf{0.9586} & 0.7376 & - \\ \midrule 
Initial $f_0$~\cite{park2022learning}       & 0.3535   & \multicolumn{1}{c|}{0.6148}& 0.9117 & 0.8044 &0.4358 \\
$f_0$ with $L_{intra}$                        & 0.5560   & \multicolumn{1}{c|}{\underline{0.6405}}& 0.9104 & 0.7676 &0.4943 \\
$f_0$ with $L_{inter}$                       & \underline{0.7051}   & \multicolumn{1}{c|}{0.6255}& 0.9547 & \underline{0.8106} & \underline{0.5321} \\
\midrule
\textbf{Ours}                               & \textbf{0.7101}   & \multicolumn{1}{c|}{\textbf{0.6635}} & \underline{0.9555} & \textbf{0.8181} & \textbf{0.5334}   \\ 
\bottomrule
\end{tabular}
}
}
\vspace*{1mm}
\end{table}

\textbf{Resnet-18}~\cite{he2016deep} is a widely used deep learning model that was pretrained on ImageNet. 
\textbf{AutoEncoder}~\cite{kramer1991nonlinear} is a deep model that learns representations by reconstructing images.
\textbf{PCA}~\cite{tipping1999probabilistic} uses a dimension reduction method. We flatten the input image and reduce its dimension to $10$ prior to analysis.
\textbf{Initial $f_0$}~\cite{park2022learning} is a scoring function that has been trained with weak supervision. It learns the order of satellite images based on annotations provided in a small sample set.
\textbf{POI}~\cite{xi2022beyond} represents an unsupervised methodology that incorporates points-of-interest (POI) within a grid image. We evaluate the image embedding produced by encoders trained through contrastive learning and apply the attentional fusion regressor, as introduced in the original literature.
\textbf{Ours} represents a complete model, including all proposed training steps.

The PCA baseline performs the worst, which is most likely due to a lack of spatial information consideration.  Following PCA, the AutoEncoder and ResNet-18, also had negative $R^{2}$ for the district population in the municipality, indicating their failure to associate human-related features with the images. The POI model had more comparable results with our models, showing slightly lower results than ours. However, the POI model utilized additional information on human activities regarding points of interests such as public transportation locations. Even without additional human information, \textbf{Ours} showed average $R^{2}$ of 0.7101 and 0.6635 for municipality and grid level respectively, which outperformed all the unsupervised baselines and performed on par or even better than the POI model. This shows that our model demonstrated robust performance at both the municipality and grid levels and across districts and countries. Because the most granular employment data is available at the municipality level, evaluations of employment size are conducted exclusively at this level.

\subsection{Ablation studies}
%\subsubsection{Ablation studies}
We compare the performance of the full model to its ablations. As shown in the lower parts of Table~\ref{tab:tbmain}, both loss objectives ${L}_{intra}$ and ${L}_{inter}$ contribute to improved prediction. ${L}_{intra}$ calibrates image score to achieve the best grid-level prediction within a district, but might loose generality across districts when used alone, as can be suggested from performance drop within a country. This can be supplemented by ${L}_{inter}$, which targets to explore an adequate score scale based on real values across district, serving as assistant for our final success. 
The observed enhancements in $R^{2}$ values for both the grid and municipality evaluations validate that our loss objectives ${L}_{intra}$ and ${L}_{inter}$ are well-aligned with our intentions.   

\begin{figure}[htpb!]
\centering
\includegraphics[width=0.5\textwidth]{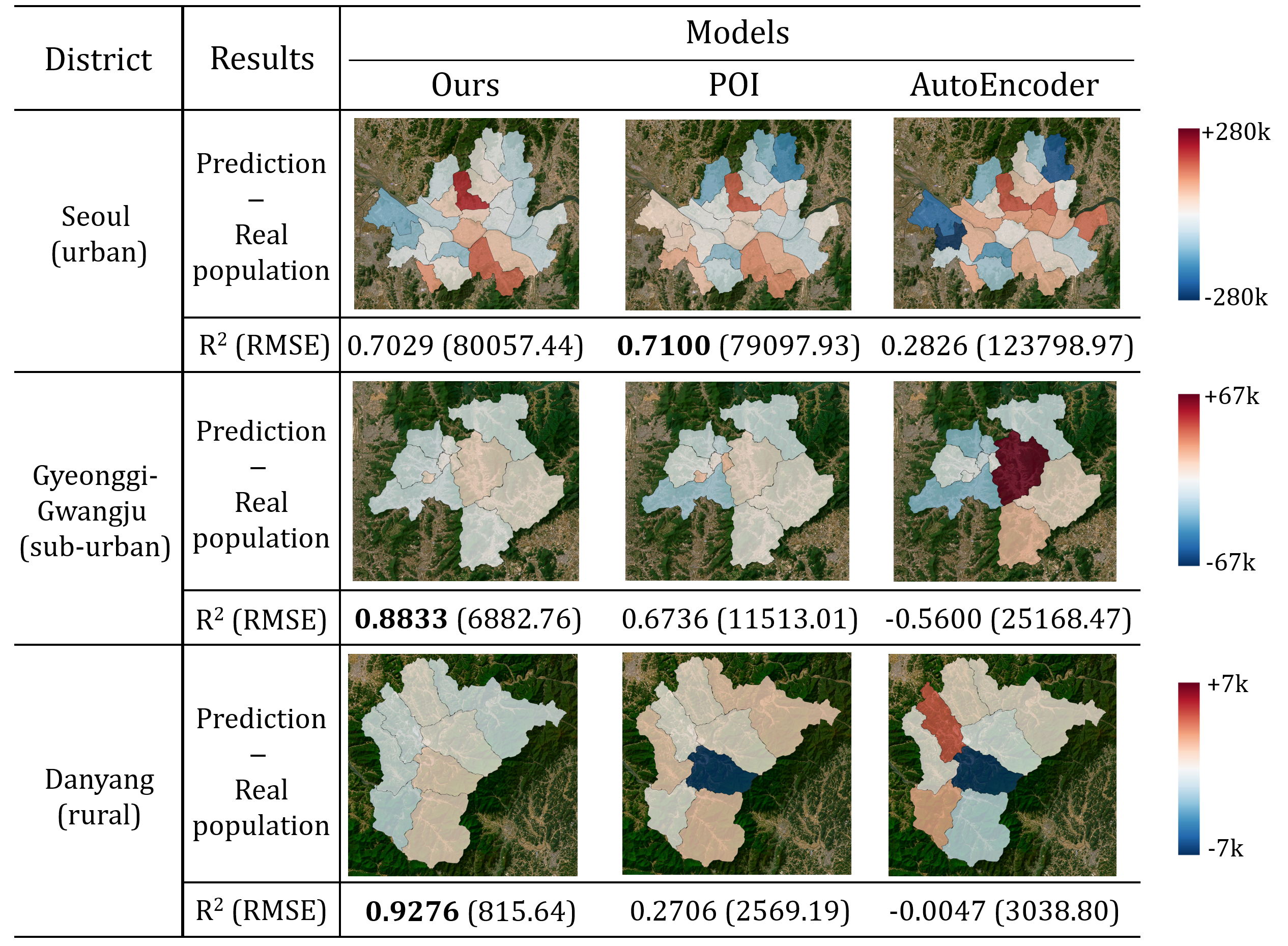} 
%\vspace{-\baselineskip}
\vspace*{-6mm}
\caption{Heatmap depicting difference between our model's prediction and the real population for three South Korean districts: Seoul (urban), Gyeonggi-Gwangju (sub-urban), and Danyang (rural). The R-squared ($R^2$) values and RMSE, tested against the ground truth population are also presented.}
\label{fig:diff_map}
\end{figure}

\section{Discussion}
Our model achieved SOTA accuracy in predicting two socioeconomic indicators in municipality and grid level.
To further explain the improvement qualitatively, we present Figure~\ref{fig:diff_map}. This demonstrates that our model's predictions are comparable to those of the existing model in urban areas like Seoul, while outperforming other models in suburban (e.g., Gyeonggi-Gwangju) and rural (e.g., Danyang) areas by exhibiting a narrower deviation from the actual population. 
The baseline models in Danyang show the greatest disparity for its municipality, a downtown hub with the most people in the district. The underestimation occurs because baseline models do not consider the relationships within a district, instead focusing on learning representations of individual images. Our model uses information about relationships within districts during training and performs well in areas with uneven development.
This finding suggests our model will be particularly effective in developing countries, where districts frequently exhibit significant disparities as a result of ongoing development.

\section{Conclusion}

This study examined alternative data sources for frequent and granular socioeconomic indicator measurement. In particular, we focused on satellite imagery, which provides periodic observations over large areas. 
We proposed a deep learning-based model that maps features observed in satellite images to socioeconomic indicators such as population and employment size at a municipality level.  
In doing so, we demonstrate that using the distribution of socioeconomic indicators inside the district and making it obey a preferred theoretical distribution (such as the common power law) is important for guiding the learning process. 
This promising method enables the utilization of distributions inferred through empirical and theoretical knowledge, such as observation or probabilistic modeling, into the learning process of satellite images. 
For example, we expect it is possible to address social issues, such as predicting temperature change in urban heat islands, by adapting a probabilistic temperature model~\cite{malings2017surface}.

%\begin{acks}

%\vspace{-0.9mm}
\vspace{1.7mm}
\noindent \textbf{Acknowledgements~} 
We thank Hyeonho Song, Sungwon Han, and Sungwon Park for their feedback. 
This work was supported by the Institute for Basic Science (IBS-R029-C2) and the National Research Foundation of Korea (NRF) grant (RS-2022-00165347). 

%\end{acks}

\onecolumn
\begin{multicols}{2}
   \bibliographystyle{ACM-Reference-Format}
   \bibliography{main}
\end{multicols}
%\bibliographystyle{ACM-Reference-Format}

%\bibliography{reference}

\end{document}